\input harvmac
\noblackbox
\newcount\figno
\figno=0
\def\fig#1#2#3{
\par\begingroup\parindent=0pt\leftskip=1cm\rightskip=1cm\parindent=0pt
\baselineskip=11pt
\global\advance\figno by 1
\midinsert
\epsfxsize=#3
\centerline{\epsfbox{#2}}
\vskip 12pt
\centerline{{\bf Figure \the\figno:} #1}\par
\endinsert\endgroup\par}
\def\figlabel#1{\xdef#1{\the\figno}}
\def\pano{\par\noindent}

\def\meno{\medskip\noindent}

\font\cmss=cmss10
\font\cmsss=cmss10 at 7pt
\def\rlx{\relax\leavevmode}
\def\inbar{\vrule height1.5ex width.4pt depth0pt}
\def\IC{\relax\,\hbox{$\inbar\kern-.3em{\rm C}$}}
\def\IN{\relax{\rm I\kern-.18em N}}
\def\IP{\relax{\rm I\kern-.18em P}}
\def\ZZ{\rlx\leavevmode\ifmmode\mathchoice{\hbox{\cmss Z\kern-.4em Z}}
 {\hbox{\cmss Z\kern-.4em Z}}{\lower.9pt\hbox{\cmsss Z\kern-.36em Z}}
 {\lower1.2pt\hbox{\cmsss Z\kern-.36em Z}}\else{\cmss Z\kern-.4em Z}\fi}
\def\narrowplus{\kern -.04truein + \kern -.03truein}
\def\narrowminus{- \kern -.04truein}
\def\narrowminussub{\kern -.02truein - \kern -.01truein}
\def\a{\alpha}
\def\b{\beta}
\def\si{\sigma}
\def\cl{\centerline}

\def\o#1{\bar{#1}}
\def\v#1{\vec{#1}}
\def\lra{\longrightarrow}
\def\rarr{\rightarrow}
\def\lb{\{ }
\def\rb{\} }

\def\qi{\vec{q}_i}
\def\qa{\vec{q}_a}
\def\z{\vec{z}}

\def\kh{K\"{a}hler }
\def\lg{Landau-Ginzburg }

\def\n{\nu}
\def\sqr#1#2{{\vcenter{\vbox{\hrule height.#2pt
            \hbox{\vrule width.#2pt height#1pt \kern#1pt
                  \vrule width.#2pt}\hrule height.#2pt}}}}
\def\square
 {\mathop{\mathchoice{\sqr{12}{15}}{\sqr{9}{12}}{\sqr{6.3}{9}}{\sqr{4.5}{9}}}}

\lref\rbw{R.\ Blumenhagen and A.\ Wi{\ss}kirchen, Nucl. Phys. {\bf B454} (1995)
561.}
\lref\rbwmod{R.\ Blumenhagen and A.\ Wi{\ss}kirchen, Nucl. Phys. {\bf B475}
(1996)
225.}
\lref\rduality{J. Distler and S. Kachru, Nucl. Phys. {\bf B442} (1995) 64.}
\lref\rbswmirror{R.\ Blumenhagen, R.\ Schimmrigk and A.\ Wi{\ss}kirchen,
 hep--th/9609167.}
\lref\rbsw{R.\ Blumenhagen, R.\ Schimmrigk and A.\ Wi{\ss}kirchen,
  Nucl. Phys. {\bf B461} (1996) 460.}
\lref\rgp{B. Greene and R. Plesser,  Nucl. Phys. {\bf B338} (1990) 15.}
\lref\rew{E.\ Witten,
  Nucl. Phys. {\bf B403} (1993) 159.}
\lref\rkm{T. Kawai and K. Mohri,
  Nucl. Phys. {\bf B425} (1994) 191.}
\lref\rdgm{J. Distler, B. Greene and D. Morrison,
  hep--th/9605222.}
\lref\rdnotes{J. Distler, ``Notes on $(0,2)$ Superconformal Field Theories'',
published in Trieste HEP Cosmology 1994, 322.}
\lref\rdk{J.\ Distler and S.\ Kachru,
  Nucl.\ Phys.\ {\bf B413} (1994) 213.}
\lref\rkw{S. Kachru and E. Witten,  Nucl.\ Phys.\ {\bf B407} (1993) 637.}
\lref\rvlg{C. Vafa, Mod. Phys. Lett. {\bf A4} (1989) 1169.}
\lref\rdiscrete{C. Vafa, Nucl. Phys. {\bf B273} (1986), 592.}
\lref\rwlg{E. Witten, Int. J. Mod. Phys. {\bf A9} (1994), 4783.}
\lref\rvi{K. Intriligator and C. Vafa, Nucl. Phys. {\bf B339} (1990), 95.}
\lref\rsw{E. Silverstein and E. Witten, Nucl. Phys. {\bf B444} (1995), 161.}
\lref\rks{M. Kreuzer and H. Skarke, Mod. Phys. Lett. {\bf A10} (1995), 1073.}
\lref\rdhvw{L. Dixon, J. Harvey, C. Vafa, and E. Witten, Nucl. Phys. {\bf B274}
(1986), 285.}
\lref\rkyy{ T. Kawai, Y. Yamada, and S-K Yang, Nucl. Phys. {\bf B414} (1994),
191.}
\lref\rvw{C. Vafa and E. Witten, J. Geom. Phys. {\bf 15} (1995), 189.}
\lref\rbconjecture{V. Batyrev, J. Alg. Geom. {\bf 3} (1994), 493.}
\lref\rsyz{A. Strominger, S-T Yau, and E. Zaslow, hep--th/9606040.}
\lref\rbh{P. Berglund and M. Henningson, Nucl. Phys. {\bf B433} (1995), 311.}

\Title{\vbox{\hbox{hep--th/9611172}
                 \hbox{IASSNS--HEP--96/120}}}
{On Orbifolds of (0,2) Models}
\smallskip
\centerline{{Ralph Blumenhagen${}^1$}  and  {Savdeep Sethi${}^2$} }
\bigskip
\centerline{${}^{1,2}$ \it School of Natural Sciences,
                       Institute for Advanced Study,}
\centerline{\it Olden Lane, Princeton NJ 08540, USA}
\smallskip
\bigskip
\bigskip\bigskip
\centerline{\bf Abstract}
\noindent
We study orbifolds of (0,2) models, including some cases with discrete torsion.
Our emphasis is on models which have a Landau-Ginzburg realization, where we
describe part of the massless spectrum by computing the elliptic genus for the
orbifolded theory. Somewhat surprisingly, we find  simple examples of (0,2)
mirror pairs that are related by a quotient action. We present a detailed
description of a family of such pairs.

\footnote{}
{\pano
${}^1$ e--mail:\ blumenha@sns.ias.edu
\pano
${}^2$ e--mail:\ sethi@sns.ias.edu
\pano}
\Date{11/96}
\newsec{Introduction}

Mirror symmetry remains one of the more mysterious, and intriguing discoveries
in
string
theory. While mirror symmetry has been intensively studied in $(2,2)$ models,
the
analogue of
mirror symmetry in $(0,2)$
models is a subject largely in its infancy. What do we mean by mirror symmetry
for
$(0,2)$ models?  Let us
recall that for $(2,2)$ models, mirror symmetry is the statement that strings
propagating on
mirror Calabi-Yau spaces, which are generally inequivalent, give equivalent
physical
theories. The superconformal field theories that describe
the infra-red physics for the mirror models are isomorphic. This isomorphism
exchanges
the moduli
corresponding to \kh deformations with those corresponding to complex structure
deformations. At the level of the N=2 superconformal algebra, this map simply
corresponds to
changing the sign of either the left or
right-moving $U(1)$ current. For $(0,2)$ models, we have an additional
set of moduli which are
holomorphic deformations of the gauge bundle. It is natural to envision
physically
equivalent
realizations of $(0,2)$ models where the role of moduli in these three classes
are
interchanged. An
example of a duality, manifest in the \lg theory, where some deformations of
the
gauge
bundle were exchanged with complex structure
deformations was described in \rduality. We shall
consider $(0,2)$ models to be mirrors if the conformal field theories to which
they
flow in the infra-red
are related by an isomorphism changing the sign of either the left or
right-moving
$U(1)$ current. This
definition, proposed and studied from a different approach in \rbswmirror, is
in
complete analogy with the $(2,2)$ case. For
compactifications on Calabi-Yau three-folds, this isomorphism exchanges the
number of
generations and
anti-generations.

The goal of this paper is to study orbifolds of $(0,2)$ models, with and
without
discrete torsion.  One of our
 aims in undertaking this study was to find examples of $(0,2)$ mirror pairs
related
by quotient actions.  We
 shall find such examples, and they are remarkably simple.  These models can be
considered $(0,2)$ analogues of the $(2,2)$ mirror pairs described by Greene
and
Plesser \rgp. The main examples we shall present flow to conformal field
theories
which have been discussed extensively in \rbw,\rbwmod\  and \rbsw. For these
models, there is reasonable evidence that the
quotient
action
that we shall describe implements the desired isomorphism on the
conformal
field theory.

In the following section, we consider the effect of orbifolding on the spectrum
of massless particles of
$(0,2)$ orbifolds.  The models that we shall consider can be obtained from the
Landau-Ginzburg phase of a gauged linear sigma model \rew. In determining the
spectrum, we shall not concern ourselves with issues of
conformal invariance, and possible non-perturbative obstructions to conformal
invariance. Our
results are only applicable to models that flow, unobstructed, to good
conformal
field theories. For example, see \rsw, for a discussion about possible
non-perturbative obstructions. To determine the spectrum, we shall generalize
the
computation
of the elliptic
genus described in \rkm.  Section three contains some models which illustrate
the
effect of orbifolding on the spectrum, including several mirror pairs. We
conclude in section four with some examples with non-trivial discrete torsion,
and a discussion of some of the issues which arise in interpreting the
quotient action in models with a Calabi-Yau phase.

\newsec{The Spectrum of $(0,2)$ Orbifolds}

\subsec{The linear sigma model}

The primary reason $(0,2)$ models have become accessible to study in recent
times is the development of the gauged linear sigma model by Witten \rew. This
model is a relatively tractable massive two-dimensional field theory which is
believed, under suitable conditions, to flow in the infra-red to a non-trivial
superconformal field theory. One of the more interesting features of the linear
sigma model is its various connected vacua, or phases. At low energies, these
phases appear to correspond to theories such as a non-linear sigma model, a
Landau-Ginzburg orbifold, or some other more peculiar theory like a hybrid
model. The linear sigma model provides a natural setting in which the relation
between some of these various types of theories can be studied.

Let us begin by describing the fields in the $(0,2)$ linear sigma model. There
are two sets of chiral superfields:  $\lb\Phi_i\vert i=1,\ldots,N_i\rb$ with
$U(1)$ charges
$\omega_i$ and $\lb P_l\vert l=1,\ldots,N_l\rb$ with $U(1)$ charges $-m_l$.
Furthermore, there are two sets of Fermi superfields: $\lb\Lambda^a\vert
a=1,\ldots,N_a\rb$
with charges $n_a$ and $\lb\Sigma^j\vert j=1,\ldots,N_j\rb$ with charges
$-d_j$.
The superpotential of the linear $\si$-model is given by,
\eqn\superpot{ S=\int d^2 z d\theta \left[ \Sigma^j W_j(\Phi_i) + P_l \Lambda^a
F_a^l(\Phi_i)
                \right], }
where $W_j$ and $F_a^l$ are quasihomogenous polynomials of degree fixed by
requiring charge neutrality of the action. To ensure the absence of gauge
anomalies, a prerequisite for conformal invariance, we demand that the
following
conditions be satisfied:

\eqn\anfree{\eqalign{  &\sum \omega_i = \sum d_j, \cr & \sum n_a = \sum m_l,
\cr
                       &\sum d_j^2 - \sum w_i^2 = \sum m_l^2 - \sum n_a^2. \cr
}}
These conditions have a simple interpretation in the Calabi-Yau phase of the
model, where the `size' of the space $r>>0$. At low energies, the model then
describes a $(0,2)$ sigma model on weighted projective space,
$\IP_{\omega_1,\ldots,\omega_{N_i}}[d_1,\ldots,d_{N_j}]$, with a coherent sheaf
of rank $N_a-N_l$ defined by the sequence:
\eqn\sheaf{  0\to\ V\to\bigoplus_{a=1}^{N_a}{\cal O}(n_a)\to
              \bigoplus_{l=1}^{N_l}{\cal O}(m_l)\to0.}
For our purposes, the sheaf can be considered a vector bundle, although the
distinction is important when resolving singularities \rdgm. The anomaly
conditions \anfree\ then correspond to the conditions for the vanishing of the
first Chern class of the tangent and vector bundles, $c_1(T)=c_1(V)=0$, and the
further topological constraint, $c_2(V)=c_2(T)$. Before proceeding, we should
comment that spectator fields are generally required in these models \rdnotes,
but since they are massive in the infra-red, we shall subsequently ignore them.

For the situation where $N_l=1$ and $r<<0$, the low-energy physics is described
by a Landau-Ginzburg orbifold. A phase which has been discussed in some detail
in \rdk.  On minimizing the scalar potential, the field $p$ gets a VEV, and the
$U(1)$ gauge symmetry is broken to a discrete $\ZZ_m$ subgroup. The dynamics of
the remaining fields is governed by a superpotential of the form,
\eqn\lgsupo{    W(\Phi_i,\Lambda_a,\Sigma_j)= \sum_j \Sigma_j W_j(\Phi_i)
                      + \sum_a \Lambda_a F_a(\Phi_i). }
For appropriate choices of the constraints $W_j$ and $F_a$, this superpotential
has an isolated singularity at the origin and is quasi-homogeneous of degree
one,
if one assigns charges $\omega_i/m$ to $\Phi_i$,
$n_a/m$ to $\Lambda_a$, and $1-d_j/m$ to $\Sigma_j$.
Quasi-homogenity implies the existence of a right-moving $R$-symmetry, and a
left-moving  $U(1)_L$. The associated currents are denoted by $J_R$ and $J_L$,
respectively.
The charges of the various fields with respect to
these $U(1)$ currents are summarized in the following table:
\vskip 0.1in
\meno
\cl{\vbox{
\hbox{\vbox{\offinterlineskip
\def\tablespace{height2pt&\omit&&\omit&&\omit&&\omit&&\omit&\cr}
\def\tablerule{\tablespace\noalign{\hrule}\tablespace}

\hrule\halign{&\vrule#&\strut\hskip0.2cm\hfil#\hfill\hskip0.2cm\cr
\tablespace
& Field && $\phi_{i}$ && $\psi_{i}$ && $\lambda_a$  &&
 $\sigma_j$  &\cr
\tablerule
& $q_L$ && ${\omega_i\over m}$ &&${\omega_i\over m}$  && ${n_a \over m}-1$
&& $-{d_j\over m}$  &\cr
\tablerule
& $q_R$ && ${\omega_i\over m}$ && ${\omega_i\over m}-1$ && ${n_a \over m}$
&& $1-{d_j \over m}$ &\cr
\tablespace}\hrule}}}}
\cl{
\hbox{{\bf Table 1:}{\it ~~Left and right charges of the
fields in the LG model.}}}
\meno
Of course, the fermions, $\psi_i$, belong to the chiral superfield, $\Phi_i$,
while the fermions, $\lambda_a$ and $\sigma_j$, are the lowest components of
the
Fermi superfields
$\Lambda_a$ and $\Sigma_j$, respectively. The techniques developed in \rkw\ can
be used to compute the
massless spectrum of such models. The simplifying feature of such calculations
is that the right-moving supersymmetry forces the massless states to lie in the
cohomology of the right-moving super charge. At the level of cohomology, the
superpotential can be set to zero, and the required calculations reduce to
those
in a free field theory. The cohomology of the charge can be
computed in a number of ways; for instance, by running a spectral sequence.

\subsec{Discrete symmetries of (0,2) LG models}

We are interested in taking further orbifolds of these $(0,2)$ Landau-Ginzburg
models, in the hope of obtaining mirror pairs. Let us simplify notation
somewhat, and consider a set of chiral
superfields, $\lb\Phi_i\vert i=1,\ldots,N\rb$, and a set of Fermi superfields
$\lb\Lambda_a\vert a=1,\ldots,M=N_a+N_j \rb$. The  superpotential is of the
form
\eqn\supo{    W(\Phi_i,\Lambda_a)= \sum_a \Lambda_a F_a(\Phi_i). }
Just as is familiar from $(2,2)$ Landau-Ginzburg theories, we can consider
discrete symmetries of the superpotential. For the sake of this discussion, we
can consider the kinetic term for these models to be the one corresponding to a
flat metric. Throughout this paper, we shall only consider symmetries which act
on the fields by phases,
\eqn\phase{ \Phi_i\rightarrow e^{2\pi i q_i} \Phi_i,
            \Lambda_a\rightarrow e^{ -2\pi i q_a} \Lambda_a. }
In particular, this excludes permutation symmetries, and non-abelian orbifolds
from our discussion.

Quotients by some of these symmetries should lead to new modular invariant
$(0,2)$ superconformal theories, and if we are fortunate, to theories whose
spectra are `mirror' to our starting model. Our starting model will always be
one obtained from a linear sigma model in the way we have just described.
Let us introduce charges, $Q^\mu$, which generate discrete abelian symmetries
of order  $h^\mu$, where the index  $\mu\in\{0,\ldots,P-1\}$. Alternatively,
$h^\mu$ is
the smallest integer such that $h^\mu q^\mu_{i,a}$ is integral for all $i$ and
$a$, where  the fields $\Phi_i$ and $\Lambda_a$ have charges $q^\mu_{i}$,
$q^\mu_{a}$
respectively  under these symmetries.
One of our orbifolds should correspond to the $\ZZ_{m}$ discrete gauge
symmetry generated in the linear sigma model. We shall set $J^0 = J_L$, the
left-moving $U(1)$ current from the linear sigma model. To obtain a string
vacuum, we shall, however, actually quotient by,
\eqn\gso{ g=e^{-\pi i Q^0} (-1)^\lambda, }
rather than $g=e^{2\pi i Q^0}$. This $\ZZ_{2m}$ symmetry is the usual GSO
projection onto integer charge states which leads to spacetime supersymmetry,
and an enhancement of the gauge group to $E_6$, $SO(10)$, or $SU(5)$ depending
on the rank of the vector bundle. The additional $\ZZ_2$ accounts for the
number
of excitations, $\lambda$, of the  left-moving fermions needed for the
$SO(16-2r)$ linearly realized gauge sector of the heterotic string. Sectors
twisted by odd powers of \gso\ are left Neveu-Schwarz, while those twisted by
even powers are left Ramond sectors.

We shall assume that the \lg theory orbifolded by \gso\ exists, and flows to a
good superconformal field theory in the infra-red. The specific examples that
we
shall consider have been realized as exact conformal field theories, and so
this
is very likely to be true. We shall then orbifold the theories further. Not all
orbifold actions result in modular invariant conformal field theories. Rather,
there exist conditions analogous to the constraints \anfree\ which had a nice
topological interpretation in the Calabi-Yau phase. For toroidal orbifolds,
such
conditions have been described by Vafa in \rdiscrete. We shall arrive at
analogous conditions for these models by studying the modular properties of the
elliptic genus. The resulting conditions are exactly of the form expected from
level matching. We shall also examine the freedom to weight the various twisted
sectors by a phase. This freedom, known as discrete torsion \rdiscrete, should
exist for $(0,2)$ models, although it has only been explored for the subclass
of
models with $(2,2)$ supersymmetry. The allowed phases are constrained if the
elliptic genus is to retain good modular properties. It seems likely that these
constraints are sufficient to guarantee modular invariance of the resulting
orbifolded theory. They are, however, certainly necessary.

\subsec{The elliptic genus of $(0,2)$ orbifolds}

Unlike $(2,2)$ models where we have A and B twisted topological theories, for
$(0,2)$ models, we only have the half-twisted model. At genus one, the
partition
function for the half-twisted model is known as the elliptic genus of the
theory.
The elliptic genus nicely encodes part of the spectrum of massless particles in
the $(0,2)$ model. For instance, the number of generations and anti-generations
can be extracted from the elliptic genus, while the remainder of the massless
spectrum requires more tedious calculations \rdk. Denoting the right-moving
R-charge by $F_R = \oint J_R$, the elliptic genus is defined by,

\eqn\genus{ Z(\tau,\n)=\Tr\, (-1)^{F_R} e^{2\pi i \n Q^0}
                                q^{L_0} e^{-\beta \o{L}_0}, }
where $q=e^{2\pi i\tau}$. It will be more convenient for us to shift $\n$ by
${1\over 2}$ which replaces $(-1)^{F_R}$ by $(-1)^F$. The right-moving N=2
supersymmetry algebra,
\eqn\susyal{ \{ \o{Q}_- , \o{Q}_+ \} = \o{L}_0,}
ensures that \genus\ is independent of $\beta$. This is the outstanding feature
of the elliptic genus, since it can then be computed in a perturbative (small
$\beta$) expansion. It further ensures that the massless states are annihilated
by $\o{L}_0$. The trace in \genus\ is taken over states in the RR sector of the
theory. Fortunately, the computation of the elliptic genus has been nicely
described for $(0,2)$ \lg theories orbifolded by the GSO projection \rkm. We
shall consider the generalized case of multiple orbifolds, and non-trivial
discrete torsion.

Let us introduce a vector notation, $\qa, \qi, $ and $ \v{Q}$ for the charges
of our discrete symmetries. It is also convenient to introduce a tensor
of charges,
\eqn\Rmatrix{ R^{\mu\nu}=\sum_{a=1}^M q^{\mu}_a q^{\nu}_a -
                        \sum_{i=1}^N q^{\mu}_i q^{\nu}_i,}
and a vector:
\eqn\rvector{ r^\mu =\sum_{a=1}^M q^\mu_a - \sum_{i=1}^N q^\mu_i. }
The elliptic genus can be expressed in terms of a Jacobi theta function. Let us
recall that,
\eqn\thetaeins{ \vartheta_1(\tau,\n)=i e^{-\pi i \n}
\sum_{n=-\infty}^{n=\infty}
(-1)^n q^{{1\over 2} (n- {1\over 2})^2} e^{2\pi i n \n},}
and that $\vartheta_1(\tau,\n)$ has the following nice modular properties,
\eqn\modular{ \eqalign{  \vartheta_1(\tau+1, \n) &= e^{\pi i \over 4}
\vartheta_1(\tau, \n) \cr
 \vartheta_1(-{1\over\tau}, {\n\over\tau}) \, &= (-i\tau)^{1\over 2} e^{\pi i
\n^2 \over \tau} \vartheta_1(\tau, \n), \cr
}}
and obeys the double quasi-periodicity relation:
\eqn\quasi{ \vartheta_1(\tau, \n+n\tau+m) = (-1)^{n+m} e^{-\pi i
(n^2\tau+2n\n)}
\vartheta_1(\tau, \n).}

To include the effect of further quotients, let us consider the elliptic genus
with the boundary condition in time twisted by the generators of the additional
discrete symmetries,
\eqn\newgenus{ Z(\tau, \n, \v{z})= \Tr \, (-1)^{F} e^{2\pi i \n Q^0} e^{2\pi i
\v{z} \v{Q}} q^{L_0} e^{-\beta \o{L}_0}, }
where $z^\mu$ is integral for all $\mu\in\{0,\ldots,P-1\}$.
Note that $\n$ is a continuous parameter
in this expression. The elliptic genus can be described in terms of the
following
function \rkm,
\eqn\untwis{ {\scriptstyle{\vec{0}}}\square_{\vec 0}\, (\tau,\n, \v{z})=
\eta(\tau)^{N-M} {\prod_a \vartheta_1(\tau,q_a^0 \n + \v{q}_a \v{z}) \over
                             \prod_i \vartheta_1(\tau, q_i^0 \n + \v{q}_i
\v{z})
}, }
where in the untwisted sector, we set $\v{z}=0$. This function is well-defined
for arbitrary values of $\v{z}$ and $\n$. The untwisted sector does not mix
with
the twisted sectors under modular transformations, but obeys the relations:
\eqn\modpropa{ \eqalign{
{\scriptstyle{\vec{0}}}\square_{\vec{0}}\,(\tau+1,\n, 0) &= e^{ \pi i
{(M-N)\over
6}}
{\scriptstyle{\vec{0}}}\square_{\vec{0}}\,(\tau,\n, 0) \cr
{\scriptstyle{\vec{0}}}\square_{\vec{0}}
\left(-{1\over \tau},{\n \over \tau},0 \right) &=e^{\pi iR^{00}
{\n^2\over\tau}}
\,
{\scriptstyle{\vec{0}}}\square_{\vec{0}}\,(\tau,\n,0)\cr
 {\scriptstyle{\vec{0}}}\square_{\vec{0}}\,(\tau,\n+\gamma\tau+\delta,0)&=
            (-1)^{r^0 (\gamma+\delta)} e^{-\pi i  R^{00} ( \gamma^2\tau
           + 2 \n \gamma)}\
            {\scriptstyle{\vec{0}}}\square_{\vec{0}}\,(\tau,\n,0), \cr  }}
where $ \gamma$ and $\delta$ are integral multiples of $h^0$. The elliptic
genus
for the sectors twisted in the time direction are easy to determine since the
twisting is generated by just inserting the charge operator into the trace:
\eqn\timetwist{ {  {\scriptstyle{\vec\beta}}\square_{\vec{0}}\,(\tau,\n, 0)=
              {\scriptstyle{\vec 0}}\square_{\vec 0}\,
               (\tau,\n, \vec\beta), \quad\quad \vec\beta\in \ZZ^P }.}
There is no arbitrariness in the way these sectors contribute to the orbifold
partition function. The remaining twisted sectors are determined, to within a
phase ambiguity, by demanding that the twisted orbifold partition function make
sense under modular transformations. For instance, the sectors twisted in space
are determined by those twisted in time by noting:
\eqn\spacetwist{\eqalign{
{\scriptstyle{\vec{0}}}\square_{\vec{\alpha}}\,(\tau,0,0) &=
 {\scriptstyle{\vec{\a}}}\square_{\vec{0}}\,(-{1\over \tau},0,0) \cr
 &= {\scriptstyle{\vec{0}}}\square_{\vec{0}}\,(-{1\over \tau},0,\v{\a}) \cr
 &= e^{\pi i \v{\a} R \v{\a} \tau}
{\scriptstyle{\vec{0}}}\square_{\vec{0}}\,(\tau,0,\tau\v{\a}). }}
There is a basic check on the $\tau$-dependent phase factor in this expression.
If we twist by the identity operator, we had better recover the expression for
the elliptic genus in the untwisted sector! Therefore, let us take $\v{\a}$ to
be
a lattice vector $\v{\gamma}$, where $\gamma^\mu$ is an integral multiple of
$h^\mu$. We then find that,
\eqn\check{\eqalign{ {\scriptstyle{\vec{0}}}\square_{\vec{\gamma}}\,(\tau,0,0)
&=
 e^{\pi i  \tau \v{\gamma} R \v{\gamma}}
{\scriptstyle{\vec{0}}}\square_{\vec{0}}\,(\tau,0,\tau  \v{\gamma}) \cr
 &= (-1)^{ \v{r} \v{\gamma}}
{\scriptstyle{\vec{0}}}\square_{\vec{0}}\,(\tau,0,0),} }
which is the desired result up to a $\tau$-independent phase factor. With
similar
motivation for the general twisted sector, we define:
\eqn\twisted{  {\scriptstyle{\vec\beta}}\square_{\vec{\alpha}}\,(\tau,\n,0)=
 e^{\pi i \v{\a} R \v{\b}} e^{\pi i(\vec\alpha R \vec\alpha\tau+
                  2\alpha^\mu R^{\mu 0} \n)} {\scriptstyle{\vec
0}}\square_{\vec
0}\,(\tau,\n, \vec\alpha\tau
                       +\vec\beta), \quad\quad \vec\alpha,\vec\beta\in \ZZ^P. }
For later convenience, we have added a phase, $e^{\pi i \v{\a} R \v{\b}},$ to
the
definition \twisted. The freedom to weight the contribution from different
twisted sectors with a phase has been explored in $(2,2)$ \lg theories
\rvi\rks,
and a mirror pair involving discrete torsion has been described in \rvw. So let
us multiply the contribution from each twisted sector by a phase, $\epsilon
(\v{\a}, \v{\b}),$ where this phase includes the freedom to add non-trivial
discrete torsion. Under modular transformations, the contribution to the
orbifold
partition function from these sectors then has the following behavior,
\eqn\modpropb{\eqalign{ \epsilon (\v{\a}, \v{\b}) \,
{\scriptstyle{\vec\beta}}\square_{\vec{\alpha}}\,(\tau+1,\n, 0) &= e^{ \pi i
{(M-N)\over 6}} \epsilon (\v{\a}, \v{\b}) \,
{\scriptstyle{\vec\alpha+\vec\beta}}\square_{\vec{\alpha}}\,(\tau,\n, 0) \cr
\epsilon (\v{\a}, \v{\b}) \,
{\scriptstyle{\vec\beta}}\square_{\vec{\alpha}}\,(-{1\over\tau},{\n\over\tau},
0)
&= e^{ \pi i R^{00} {\n^2\over \tau}} \epsilon (\v{\a}, \v{\b}) \, \,
{\scriptstyle{-\v{\a}}}\square_{\vec{\b}}\,(\tau,\n, 0). \cr}}
Sectors related by modular transformations should appear with the same phase in
the orbifold partition function, which implies that,
\eqn\dismoda{\eqalign{
\epsilon(\vec\alpha,\vec\beta)&=\epsilon(\vec\alpha,\vec\alpha+\vec\beta) \cr
        \epsilon(\vec\alpha,\vec\beta)&=\epsilon(\vec\beta,-\vec\alpha). \cr }}
We can now consider a consistency requirement, which is essentially level
matching. We demand that shifting the twist operators in the space and time
directions by the identity operator leave the elliptic genus for the twisted
sector invariant,
\eqn\levmatch{ \epsilon (\v{\a}+\v{\gamma}, \v{\b}+\v{\delta}) \,
{\scriptstyle{\vec\beta+ \v{\delta} }}{\square_{\vec\alpha+ \v{\gamma} }} \,
                  (\tau,\n, 0)=\epsilon (\v{\a}, \v{\b}) \,
{\scriptstyle{\vec\beta\,}}{\square_{\vec\alpha}} \,
                 (\tau,\n, 0), }
where $\gamma^\mu, \delta^\mu$ are integral multiples of $h^\mu$. This
condition
implies the following constraint on the phases,
\eqn\newdismoda{
   \epsilon (\v{\a}+\v{\gamma}, \v{\b}+\v{\delta}) \, = \epsilon (\v{\a},
\v{\b})
\, e^{-\pi i( \v{r} \{ \v{\gamma} + \v{\delta} \} - \v{\gamma} R \v{\delta} +
\v{\gamma} R \v{\b} - \v{\a} R \v{\delta} ). }}
This condition provides us with strong constraints on the possible charge
vectors.
We shall take the phase factors to be of the general form:
\eqn\dismodb{ \epsilon(\vec\alpha,\vec\beta)=e^{\pi i \vec{w}(\vec\alpha +
              \vec\beta)}\,  e^{\pi i  \vec\alpha Q \vec\beta }. }
The term $ e^{\pi i  \vec\alpha Q \vec\beta }$ automatically satisfies the
factorization constraint which arises at genus two \rdiscrete. These conditions
imply that $w^\mu$ must be integral, and that the following conditions must be
met:
\eqn\dismodc{\eqalign{
Q^{\mu\n} + Q^{\n\mu} \in 2\ZZ, \cr
w^\mu + Q^{\mu\mu} \in 2\ZZ, \cr
(w^\mu - r^\mu)h^\mu = 0 \quad {\rm mod}\ 2, \cr
(Q^{\mu\n} + R^{\mu\n}) h^\n = 0 \quad {\rm mod}\ 2, \cr
}}
for any $\mu,\nu\in\{0,\ldots,P-1\}$. In turn, these conditions provide
constraints on $r^\mu$ and $R^{\mu\n}$, depending on whether $h^\mu$ is even:
\eqn\even{ \eqalign{ r^\mu h^\mu \in 2\ZZ, \cr R^{\mu\mu} h^\mu \in 2\ZZ,}}
or odd,
\eqn\odd{ \eqalign{ r^\mu h^\mu \in \ZZ, \cr R^{\mu\mu} h^\mu \in \ZZ.}}
These conditions are very similar to those described in \rdhvw. The final set
of
constraints we need to consider relate $r^\mu$ to $R^{0\mu}$. In the case of a
single quotient, the condition,
\eqn\spectralconst{ r^0 = R^{00},}
can be seen to arise by requiring reasonable behavior of the twisted partition
function under the spectral flow acting on the left $U(1)$ current, $J^0$, and
the
energy-momentum tensor $T$ \rkyy.
Condition \spectralconst\ can also be seen to arise from the requirement
that the anomalies in the operator product expansion (OPE) of $T$ and $J_R$
with
$J^0$ vanish.
To study the general case of multiple quotients, let us momentarily
restrict to those discrete symmetries which can be promoted to continuous
symmetries of the theory. In this case, the superpotential satisfies,
\eqn\contsymm{ W(e^{-2\pi i \theta q_a^\mu} \Lambda_a, e^{2\pi i \theta
q_i^\mu} \Phi_i) = W(
\Lambda_a, \Phi_i), }
for any $ \theta$. Let us further restrict our discussion to the level of
$\o{Q}_+$-cohomology \rwlg, where we can define a $U(1)$ current for this
symmetry:
\eqn\currents{ J^\mu =  - \sum q_a^\mu \lambda_a {\bar \lambda}_a - \sum
q_i^\mu (\partial_-
\phi_{\bar \iota} ) \phi_i, }
which is well-defined in $\o{Q}_+$-cohomology. Requiring that the OPE between
$T$
and $J^\mu$ be non-anomalous yields,
\eqn\opecond{ r^\mu = R^{0\mu}.}
However, since we are actually considering only discrete symmetries, it seems
reasonable to impose the weaker condition:
\eqn\finalconst{ (r^\mu - R^{0\mu}) h^\mu \in \ZZ,}
together with $ r^0=R^{00}$. We shall restrict our discussion to orbifolds of
models which satisfy the constraint \finalconst. Finally, we can define the
orbifold partition function,
\eqn\orbieg{ Z_{orb}(\tau,\n)={1\over \prod{ h^\mu}}
       \sum_{\alpha^0,\beta^0=0}^{h^0-1}
       \ldots \sum_{\alpha^{P-1},\beta^{P-1}=0}^{h^{P-1}-1} \, \epsilon(\v{\a},
\v{\b}) \, \,
         {\scriptstyle{\vec\beta\,}}{ \square_{\vec\alpha} }\,(\tau,\n, 0), }
which transforms under modular transformations according to \modpropa.

In order to obtain  information about the massless spectra, we need only study
the leading terms in an expansion of \orbieg\ in powers of $q$. Let $y=e^{2\pi
i\n}$, then, following the definition in \rkm, the $\chi_y$ genus is given by,
\eqn\chiy{ \chi_y = \lim_{q\rightarrow 0} \, (i)^{N-M} q^{ N-M\over 12}
y^{{1\over 2} r^0} Z_{orb}(q,y). }
We shall denote the contribution to $\chi_y$ from a twisted sector $\v{\a}$ by
$\chi^{\v{\a}}_y$. The contribution from each twisted sector is determined in
terms of the function,

\eqn\chiyorb{ f^{\v\alpha}(\vec{z}) =(-1)^{\vec{w}\vec\alpha} e^{2\pi i
\z\vec{Q}_{\vec\alpha}}
                               q^{E_{\vec\alpha}}
                    {\prod_a (-1)^{[\vec\alpha\qa]}
      (1 - e^{2\pi i \z\qa} q^{\lb\vec\alpha\qa\rb} )
      (1 - e^{-2\pi i \z\qa} q^{1-\lb\vec\alpha\qa\rb} ) \over
      \prod_i (-1)^{[\vec\alpha\qi]}
      (1- e^{2\pi i \z\qi} q^{\lb\vec\alpha\qi\rb} )
      (1- e^{-2\pi i \z\qi} q^{1-\lb\vec\alpha\qi\rb} ) }, }
where $\chi^{\v{\a}}_y$ is given by expanding $f^{\v{\a}} (\v{z})$ in powers of
$q$,
and retaining terms of the form $q^0 e^{-2\pi i \z(\vec\sigma+\vec{n})}$,
where $\v{n}\in\ZZ^P$ and $\v\sigma = {1\over 2} \v{w} + {1\over 2} \v{\a} (Q -
R)$.
Finally, we set $z_1=\ldots=z_{P-1}=0$.
Furthermore,  we have used the abreviation $\{x\}=x-[x]$ in \chiyorb.
The fractionalized
charges and energies in the twisted sectors are given by the formulae:
\eqn\charges{ \eqalign{ \v{Q}_{\vec\alpha}&=\sum_a \v{q}_a
(\vec\alpha\qa-[\vec\alpha\qa]-{1\over 2})-
                             \sum_i \v{q}_i
(\vec\alpha\qi-[\vec\alpha\qi]-{1\over
2}) \cr
       E_{\vec\alpha}&={1\over 2}\sum_a (\vec\alpha\qa-[\vec\alpha\qa]-1)
                    (\vec\alpha\qa-[\vec\alpha\qa])
                      -{1\over 2}\sum_i (\vec\alpha\qi-[\vec\alpha\qi]-1)
                      (\vec\alpha\qi-[\vec\alpha\qi]).\cr}}
To serve as the internal sector of a heterotic string compactification, we need
to impose further constraints on the allowed models. Firstly, to combine the
internal sector of the theory with the space-time and gauge sectors, we require
that in all the twisted sectors, $Q_{L\v\a}-Q_{R\v\a}\in\ZZ$, where we have
denoted
the right-moving charge in the twisted sector $\v{\a}$ by $Q_{ R\v{\alpha}}$.
We find that:
\eqn\chargeright{Q_{ R\v{\alpha}}=\sum_a q_{R
a}(\vec\alpha\qa-[\vec\alpha\qa]-{1\over 2})- \sum_i q_{R i}
(\vec\alpha\qi-[\vec\alpha\qi]-{1\over 2}), }
which implies that,
\eqn\conda{ \sum_a \qa - \sum_i \qi \in \ZZ^P.}
In addition, we require that our orbifold actions preserve the vacuum from the
untwisted sector. Among the problems that occur if this state is projected out
is
the lack of enough gauginos to obtain an $E_6$, $SO(10)$ or $SU(5)$ enhanced
gauge group. To preserve the vacuum, we require that
\eqn\condb{ \vec{w}=\left( \sum_a \qa - \sum_i \qi \right)\quad {\rm mod}\ 2.}
Lastly, we want our canonical projection onto states with left-moving charge
$q_L={1\over 2} r^0 \, {\rm mod}\ \ZZ$. This requirement leads to the
condition,
\eqn\condc{ \sum_{\mu} \alpha^\mu
                       (Q^{\mu 0}-R^{\mu 0}) \in 2\ZZ }
for every twisted sector, $\vec\alpha$. This determines $Q^{\mu 0}$ in terms of
$R^{\mu 0}$ mod 2.

\newsec{Mirror Pairs}

Our first example is the (0,2) model with the following field content, and
$U(1)$
charge assignments.
\meno
\cl{\vbox{
\hbox{\vbox{\offinterlineskip
\def\tablespace{height2pt&\omit&&\omit&&\omit&&\omit&\cr}
\def\tablerule{\tablespace\noalign{\hrule}\tablespace}

\hrule\halign{&\vrule#&\strut\hskip0.2cm\hfil#\hfill\hskip0.2cm\cr
\tablespace
& Field && $\Phi^{1,2,3,4}$ && $\Phi^{5,6}$  &&
 $\lambda^{1,\ldots,7}$  &\cr
\tablerule
& $q_l$ && ${1\over 5}$ && ${2\over 5}$ && $-{4\over 5}$  &\cr
\tablerule
& $q_r$ && ${1\over 5}$ && ${2\over 5}$ && ${1\over 5}$ &\cr
\tablespace}\hrule}}}}
\cl{
\hbox{{\bf Table 2:}{\it ~~Left and right charges of the
fields in the LG theory.}}}
\meno
This model has been studied in previous work \rbsw. The relevant geometry in
the
Calabi-Yau phase corresponds to the vector-bundle defined by the exact
sequence,
\eqn\sequence{0\to\ V\to\bigoplus_{a=1}^{5}{\cal O}(1)\to{\cal O}(5)\to0,}
over the threefold configuration $\IP_{(1,1,1,1,2,2)}[4,4]$.
In the \lg phase, the massless sector   contains  $N_{16}=80$ chiral
multiplets which transform in the
spinor representation of $SO(10)$. There are no states transforming in the
conjugate spinor representation, and there are $N_{10}=74$ chiral multiplets
which transform in the vector representation.
The description of the conformal field theory for this model is known,
and the exactly solvable point corresponds to a \lg theory with superpotential
\rbwmod,
\eqn\superpot{ W=\sum_{i=1}^4 \Lambda_i \Phi_i^4 +  \Lambda_5 \Phi_5^2
                  + \Lambda_6 \Phi_6^2 +\Lambda_7 \Phi_5 \Phi_6.}
Let us begin with some general observations.
Clearly, the superpotential is invariant under the discrete symmetry given by:
\eqn\gensymm{\eqalign{ &\phi_1\rarr e^{{2\pi i \over n} } \phi_1, \quad\quad
             \phi_2\rarr e^{-{2\pi i \over n} } \phi_2 \cr
                &\lambda_1\rarr e^{-2\pi i {4\over n} } \lambda_1, \quad\quad
             \lambda_2 \rarr e^{2\pi i {4\over n} }  \lambda_2 .\cr }}
The charges of the various fields with respect to this new symmetry are
therefore,
\eqn\gencharge{ (\vec{q}_i;\vec{q}_a)=\left({1\over n},-{1\over n},0,0,0,0;
            {4\over n},-{4\over n},0,0,0,0,0\right). }
In the subsequent discussion,
we will use this shorthand notation to denote the action of the discrete
symmetries on the fields. Calculating $r^\mu$ and $R^{\mu\nu}$ gives,
\eqn\dataa{ \vec{r}=(4,0), \quad\quad R=\left( \matrix{ 4 & 0 \cr 0 & {30\over
n^2} \cr}
\right).}
The choice of $n$ is strongly constrained  by the conditions \even\ and \odd,
leaving
only $n=3,5,15$ as allowed possibilities. This example demonstrates how
restrictive the previous conditions really are. Note, that unlike the
corresponding
$(2,2)$ quintic, we are not forced to consider  $\ZZ_5$ actions only.

Now, let us consider a $\ZZ_5$ discrete symmetry, which we  denote $G_{1,2}$,
that is given by:
\eqn\symma{\eqalign{ &\phi_1\rarr e^{{2\pi i \over 5} } \phi_1, \quad\quad
             \phi_2\rarr e^{-{2\pi i \over 5} } \phi_2 \cr
                &\lambda_1\rarr e^{{2\pi i \over 5} } \lambda_1, \quad\quad
             \lambda_2 \rarr e^{-{2\pi i \over 5} }  \lambda_2 .\cr }}
The charges of the various fields with respect to this new symmetry are
therefore,
\eqn\chargea{ (\vec{q}_i;\vec{q}_a)=\left({1\over 5},-{1\over 5},0,0,0,0;
            -{1\over 5},{1\over 5},0,0,0,0,0\right), }
where,
\eqn\dataa{ \vec{r}=(4,0), \quad\quad R=\left( \matrix{ 4 & 0 \cr 0 & 0 \cr}
\right).}
This choice of charges satisfies the constraints described in the
previous section with trivial discrete torsion. All the examples in this
section
will have trivial discrete torsion.

The result of applying the general formula for the $\chi_y$ genus from the
previous section to this example is summarized in the following table. The
table
is arranged so that the $(l,a)$ entry gives the $\chi_y$ genus for the $(l,a)$
twisted sector, where $l$ and $a$ denote the twisted sectors of the
$G_{1,2}$ quotient and the GSO quotient, respectively.

\meno
\cl{\vbox{
\hbox{\vbox{\offinterlineskip
\def\tablespace{height2pt&\omit&&\omit&&\omit&&\omit&&\omit&&\omit&\cr}
\def\tablerule{\tablespace\noalign{\hrule}\tablespace}

\hrule\halign{&\vrule#&\strut\hskip0.2cm\hfil#\hfill\hskip0.2cm\cr
\tablespace
& $l\backslash a$ && $0$ && $1$ && $2$  && $3$ && $4$  &\cr
\tablerule
& $0$ && $1+20(y-y^3)-y^4$ && $y^4$ && $-2y^2$  && $2y^2$
&& $-1$  &\cr
\tablerule
& $1$ && $5(y-y^3)$ && $0$ && $-2y^2$  && $2y^2$
&& $0$  &\cr
\tablerule
& $2$ && $5(y-y^3)$ && $y^3$ && $y^2-y^3$  && $y-y^2$
&& $-y$  &\cr
\tablerule
& $3$ && $5(y-y^3)$ && $y^3$ && $y^2-y^3$  && $y-y^2$
&& $-y$  &\cr
\tablerule
& $4$ && $5(y-y^3)$ && $0$ && $-2y^2$  && $2y^2$
&& $0$  &\cr
\tablerule}\hrule}}}}
\cl{
\hbox{{\bf Table 3:}{ ~~ ${\chi_y}$ \it genus for $G_{1,2}$. }}}
\meno
Summing up the contributions from all the twisted sectors gives the result:
\eqn\totalgenus{  \chi_y=40(y-y^3).}
To obtain the number of generations and anti-generations, we only have to count
the positive and negative $y^1$ terms in table 3 separately.
Therefore, we see that there are $N_{16}=42$ generations and
$\o{N}_{{16}}=2$ anti-generations in this model. The number of ${\bf 10}$'s
that
arise from the $a=0$ sectors can not be extracted from the $\chi_y$ genus.
Rather, a detailed analysis of the cohomology is needed to reveal that
$(a,l)=(0,0)$ contains $N_{10}=18$ vectors, and each of the other $(a,l)=(0,l)$
sectors contain $N_{10}=4$ vectors. Thus, we end up with $N_{10}=42$ vectors.
As expected from the discussion in \rbswmirror, this model has the same
spectrum
as the (0,2) model defined by the following bundle and Calabi-Yau data,
\eqn\modela{ V(3,4,4,4,5;20)\lra \IP_{3,4,4,5,8,8}[16,16]. }
This data should be viewed as input information for a $(0,2)$ linear sigma
model,
whose Calabi-Yau phase is described by this geometric information.

We can carry out a similar analysis by quotienting our starting model by
another
$\ZZ_5$ action, denoted $G_{2,3}$, given by:
\eqn\chargeb{ (\vec{q}_i;\vec{q}_a)=\left(0,{1\over 5},-{1\over 5},0,0,0;
            0,-{1\over 5},{1\over 5},0,0,0,0\right). }
The resulting spectrum then agrees exactly with the model described by,
\eqn\modelb{ V(13,15,16,16,20;80)\lra \IP_{13,15,16,20,32,32}[64,64], }
with $(N_{16},{\o{N}}_{{16}})=(8,36)$. According to the mirror construction
described in \rbswmirror,  we might have expected this model
to be the mirror of our first example where we quotiented by $G_{1,2}$.
However,
this is not the case, and we are left with the question: where are the models
with $(N_{16},{\o{N}}_{{16}})=(2,42)$ and $(N_{16},{\o{N}}_{{16}})=(36,8)$,
respectively?

A little trial and error reveals that after dividing our starting model by a
$\ZZ_5$, denoted $H_1$, which acts according to,
\eqn\chargec{ (\vec{q}_i;\vec{q}_a)=\left(-{4\over 5},0,0,0,{3\over 5},{3\over
5};
            {4\over 5},0,0,0,-{4\over 5},{1\over 5},{1\over 5}\right), }
we get the following contributions from the twisted sectors to the
$\chi_y$ genus.
\meno
\cl{\vbox{
\hbox{\vbox{\offinterlineskip
\def\tablespace{height2pt&\omit&&\omit&&\omit&&\omit&&\omit&&\omit&\cr}
\def\tablerule{\tablespace\noalign{\hrule}\tablespace}

\hrule\halign{&\vrule#&\strut\hskip0.2cm\hfil#\hfill\hskip0.2cm\cr
\tablespace
& $l\backslash a$ && $0$  && $1$ && $2$
&& $3$ && $4$   &\cr
\tablerule
& $0$ && $1+18(y-y^3)-y^4$ && $y^4$ && $-2y^2$  && $2y^2$
&& $-1$  &\cr
\tablerule
& $1$ && $6y+6y^2$ && $0$ && $y^3$  && $-2y$
&& $-3y+3y^2$  &\cr
\tablerule
& $2$ && $-12y^2-12y^3$ && $-y^2$ && $0$  && $0$
&& $-2y$  &\cr
\tablerule
& $3$ && $12y+12y^2$ && $2y^3$ && $0$  && $0$
&& $y^2$  &\cr
\tablerule
& $4$ && $-6y^2-6y^3$ && $-3y^2+3y^3$ && $2y^3$  && $-y$
&& $0$  &\cr
\tablerule}\hrule}}}}
\cl{
\hbox{{\bf Table 4:}{ ~~ ${\chi_y}$ \it genus for $H_{1}$. }}}
\meno
Again summing the twisted sector contributions gives, $\chi_y=28(y-y^3)$, and
the
desired spectrum, $(N_{16},\o{N}_{{16}})=(36,8)$, of generations and
anti-generations.
Acting with a second orbifold action, $H_2$,
\eqn\charged{ (\vec{q}_i;\vec{q}_a)=\left(0,-{4\over 5},0,0,{3\over 5},{3\over
5};
            0,{4\over 5},0,0,-{4\over 5},{1\over 5},{1\over 5}\right), }
we obtain a model with spectrum $(N_{16},\o{N}_{{16}})=(2,42)$.
We can combine the two symmetries $G_{1,2}$ and $H_3$ into a single action
$(GH)_{1,2,3}$:
\eqn\charged{ (\vec{q}_i;\vec{q}_a)=\left({1\over 5},-{1\over 5},-{4\over 5},
            0,{3\over 5},{3\over 5};-{1\over 5},{1\over 5},
            {4\over 5},0,-{4\over 5},{1\over 5},{1\over 5}\right). }
In this case, the $\chi_y$ genus for each twisted sector is displayed in the
following table.
\meno
\cl{\vbox{
\hbox{\vbox{\offinterlineskip
\def\tablespace{height2pt&\omit&&\omit&&\omit&&\omit&&\omit&&\omit&\cr}
\def\tablerule{\tablespace\noalign{\hrule}\tablespace}

\hrule\halign{&\vrule#&\strut\hskip0.2cm\hfil#\hfill\hskip0.2cm\cr
\tablespace
& $l\backslash a$ && $0$  && $1$ && $2$
&& $3$ && $4$   &\cr
\tablerule
& $0$ && $1+15(y-y^3)-y^4$ && $y^4$ && $-2y^2$  && $2y^2$
&& $-1$  &\cr
\tablerule
& $1$ && $0$ && $-3y^2+3y^3$ && $y^3$ && $-y$
&& $-3y+3y^2$  &\cr
\tablerule
& $2$ && $0$ && $-y$ && $-3y+3y^2$ && $-3y^2+3y^3$
&& $y^3$  &\cr
\tablerule
& $3$ && $0$ && $2y^2$ && $-4y$  && $4y^3$
&& $-2y^2$  &\cr
\tablerule
& $4$ && $0$ && $y-7y^2-2y^3$ && $2y^3$  && $-2y$
&& $2y+7y^2-y^3$  &\cr
\tablerule}\hrule}}}}
\cl{
\hbox{{\bf Table 5:}{ ~~${\chi_y}$ \it genus for $(GH)_{1,2,3}$. }}}
This model has $(N_{16}, \o{N}_{{16}})=(18,14)$. To find a candidate mirror, we
must quotient by $G_{1,2}\times H_3$. Finally, the mirror of our original model
is given by the quotient $G_{1,2}\times G_{2,3}\times G_{3,4}$, which gives a
model with massless spectrum: $(N_{16},\o{N}_{{16}})=(0,80)$\footnote{$^1$}{We
thank
M. Flohr for helping us with a C-code program to calculate all 625 twisted
sectors
of this model}.
To summarize, starting from the $(0,2)$ model described in table 2, we have, by
successive orbifolds,  obtained the following mirror symmetric family of
models:
\meno
\cl{\vbox{
\hbox{\vbox{\offinterlineskip
\def\tablespace{height2pt&\omit&&\omit&&\omit&\cr}
\def\tablerule{\tablespace\noalign{\hrule}\tablespace}

\hrule\halign{&\vrule#&\strut\hskip0.2cm\hfil#\hfill\hskip0.2cm\cr
\tablespace
& $G$ && $N_{16}$ && $N_{\o{16}}$  &\cr
\tablerule
& $1$ && $80$ && $0$  &\cr
\tablerule
& $G_{1,2}$ && $42$ && $2$  &\cr
\tablerule
& $H_1$   && $36$ && $8$  &\cr
\tablerule
& $(GH)_{1,2,3}$   && $18$ && $14$  &\cr
\tablerule
& $G_{1,2}\times H_3$   && $14$ && $18$  &\cr
\tablerule
& $G_{1,2}\times G_{2,3}$ && $8$ && $36$  &\cr
\tablerule
& $H_1\times H_2$   && $2$ && $42$  &\cr
\tablerule
& $G_{1,2}\times G_{2,3}\times G_{3,4}$ && $0$ && $80$  &\cr
\tablerule}\hrule}}}}
\cl{
\hbox{{\bf Table 6:}{\it ~~ A family of (0,2) models obtained by
orbifolding.}}}
\meno
In many ways, this appears to be the $(0,2)$ analogue of the Greene-Plesser
construction
for the mirror of a Fermat type $(2,2)$ theory \rgp. In the following section,
we
shall consider models with non-trivial discrete torsion, where we will find a
similar pattern.

\newsec{Examples with Non-trivial Discrete Torsion}
\meno
In order to include discrete torsion, we need to choose a finite group $G$ such
that $H^2(G,U(1))$ is non-trivial. Perhaps the simplest choice is to take $G$
to
be the product of two simple abelian groups. For example, since
$H^2(\ZZ_5\times
\ZZ_5,U(1))=\ZZ_5$, the models obtained by orbifolding by $G_{1,2}\times
G_{2,3}$, $H_1\times H_2$, and $G_{1,2}\times H_3$ admit the introduction of
a non-trivial discrete torsion.
\eqn\distora{\eqalign{
    \epsilon(\vec\alpha,\vec\beta)&=e^{\pi i (\vec\alpha Q \vec\beta) } \cr
    Q&=\left( \matrix{ 4 & 0 & 0 \cr
                0 & 0 & {2m\over 5} \cr
                0 & -{2m\over 5} & 0 \cr}\right)\quad\quad m\in\{0,\ldots,4\}
\cr }}
Including the discrete torsion has the effect of changing the projections in
the
various twisted sectors. After a tedious calculation, we arrive at the
following
spectra for these models:
\meno
\cl{\vbox{
\hbox{\vbox{\offinterlineskip
\def\tablespace{height2pt&\omit&&\omit&&\omit&&\omit&&\omit&&\omit&\cr}
\def\tablerule{\tablespace\noalign{\hrule}\tablespace}

\hrule\halign{&\vrule#&\strut\hskip0.2cm\hfil#\hfill\hskip0.2cm\cr
\tablespace
& $m$ && $0$ && $1$ && $2$ && $3$ && $4$   &\cr
\tablerule
& $G_{1,2}\times G_{2,3}$ && $(8,36)$ && $(23,3)$ && $(23,3)$ && $(23,3)$
&& $(23,3)$   &\cr
\tablerule
& $H_1\times H_2$ && $(2,42)$ && $(17,9)$ && $(19,11)$ && $(19,11)$
&& $(17,9)$   &\cr
\tablerule
& $G_{1,2}\times H_3$ && $(14,18)$ && $(16,8)$ && $(14,6)$ && $(14,6)$
&& $(16,8)$   &\cr
\tablerule}\hrule}}}}
\cl{
\hbox{{\bf Table 7:}{ ~~$(N_{16},\o{N}_{{16}})$ \it with the inclusion of
discrete
torsion.}}}
For $G_{1,2}\times G_{2,3}\times G_{3,4}$, we shall consider the following
choices for discrete torsion, which are captured in the following $Q$ matrix,
\eqn\distorb{\eqalign{
    \epsilon(\vec\alpha,\vec\beta)&=e^{\pi i (\vec\alpha Q \vec\beta) } \cr
    Q&=\left( \matrix{ 4 & 0 & 0 & 0\cr
                0 & 0 & {2m_1\over 5} & -{2m_2\over 5} \cr
                0 & -{2m_1\over 5} & 0 & {2m_3\over 5} \cr
                0 & {2m_2\over 5} & -{2m_3\over 5} & 0 \cr}\right)
                \quad\quad m_i\in\{0,\ldots,4\}.  \cr }}
Because of the permutation symmetry, there are redundant choices for the three
parameters $m_i$. Taking permutations into account leaves nineteen independent
triples. However, only six of the resulting models have different spectra, and
these cases are displayed below.

\meno
\cl{\vbox{
\hbox{\vbox{\offinterlineskip
\def\tablespace{height2pt&\omit&&\omit&&\omit&&\omit&&\omit&&\omit&&\omit&\cr}
\def\tablerule{\tablespace\noalign{\hrule}\tablespace}

\hrule\halign{&\vrule#&\strut\hskip0.2cm\hfil#\hfill\hskip0.2cm\cr
\tablespace
& $(m_1,m_2,m_3)$ && $(0,0,0)$ && $(1,1,1)$ && $(2,2,2)$ && $(1,2,2)$ &&
 $(1,3,3)$ && (1,2,3)   &\cr
\tablerule
& $(N_{16},N_{\o{16}})$ && $(0,80)$ && $(9,17)$ && $(11,19)$ && $(8,16)$
&& $(6,14)$ && (3,23)   &\cr
\tablerule}\hrule}}}}
\cl{
\hbox{{\bf Table 8:}{\it ~~ The spectra of the $G_{1,2}\times G_{2,3}\times
G_{3,4}$ orbifold with discrete torsion. }}}
The spectra for these models again form a mirror symmetric set, even with the
inclusion of discrete torsion.

If $(2,2)$ mirror symmetry is any indicator, we have only uncovered a small
part
of what is likely to be a very rich area to study. There are many avenues to
explore: perhaps, the most natural question is whether a reasonable
combinatoric
conjecture for $(0,2)$ mirror pairs can be formulated, analogous to Batyrev's
conjecture for Calabi-Yau spaces constructed from toric varieties
\rbconjecture.
In a similar vein, could a T-duality be responsible for $(0,2)$ mirror symmetry
\rsyz?
There seem to be many interesting subtleties in even trying to define, in the
Calabi-Yau phase, the operation which corresponds to the orbifold action in the
\lg phase. This is a topic that deserves detailed study. In the cases where we
have corresponding geometric models, there are numerous more mathematical
questions to answer: for instance,
desingularizing the corresponding geometric models, and checking that their
spectra agree with the \lg computations. A topic barely explored in $(2,2)$
mirror symmetry is the question of
non-abelian quotients. It would be interesting to extend this analysis to
encorporate non-abelian orbifolds.

Constructing more examples of the kind we have presented here would also be
interesting. Perhaps, a more systematic study of the properties of the elliptic
genus for this class of models would provide insight into the general structure
of $(0,2)$ mirror symmetry, as in \rbh. Further, it does not seem unlikely that
some deformation theoretic couplings are immune to instanton corrections. Can
we
use $(0,2)$ mirror symmetry to check this possibility? As a final comment, we
suspect that $(0,2)$ mirror pairs will be very helpful in unraveling part of
the
structure of N=1 dualities in four dimensions. For such applications, it is
interesting to consider $(0,2)$ models on base spaces which are
elliptically-fibered. Imposing the condition that the base admit an
elliptic-fibration undoubtedly provides interesting constraints on the allowed
bundle structure for these theories.

\bigbreak\bigskip\bigskip\centerline{{\bf Acknowledgements}}\nobreak

It is a pleasure to thank M. Flohr and D. Morrison for helpful discussions. The
work of R.B. is supported by NSF grant PHY--9513835, while that of S.S. is
supported by NSF grant DMS--9627351.
\vfill\eject

\listrefs
\bye